\documentclass[10pt,english,aps,showpacs,letter,longbibliography, twocolumn]{revtex4-1}
\usepackage{amsmath}
\usepackage[    bookmarks,
                 bookmarksopen = true,
                 bookmarksnumbered = true,
                 linktocpage,
                 colorlinks = true,
                 linkcolor = blue,
                 urlcolor  = blue,
                 citecolor = magenta,
                 anchorcolor = green,
                 hyperindex = true,
                 hyperfigures]
        {hyperref}

\usepackage[T1]{fontenc}
\usepackage[utf8]{inputenc}
\usepackage{enumitem}
\usepackage{graphicx}
\usepackage{esint}
\usepackage[utf8]{inputenc}
\usepackage[T1]{fontenc}
\usepackage{subcaption}

\usepackage{listings}
\DeclareFixedFont{\ttb}{T1}{txtt}{bx}{n}{8} % for bold
\DeclareFixedFont{\ttm}{T1}{txtt}{m}{n}{8}  % for normal

\usepackage{color}
\definecolor{deepblue}{rgb}{0,0,0.5}
\definecolor{deepred}{rgb}{0.6,0,0}
\definecolor{deepgreen}{rgb}{0,0.5,0}

\newcommand\pythonstyle{\lstset{
language=Python,
basicstyle=\ttm,
otherkeywords={self},             % Add keywords here
keywordstyle=\ttb\color{deepblue},
emph={MyClass,__init__},          % Custom highlighting
emphstyle=\ttb\color{deepred},    % Custom highlighting style
stringstyle=\color{deepgreen},
frame=tb,                         % Any extra options here
showstringspaces=false            % 
}}

\lstnewenvironment{python}[1][]
{
\pythonstyle
\lstset{#1}
}
{}

\makeatletter

\usepackage{babel}

\makeatother

\begin{document}

\preprint{This line only printed with preprint option}

%%%%% online xyscan:  http://arohatgi.info/WebPlotDigitizer/app/?

\title{Interpretable deep learning for nuclear deformation in heavy ion collisions}

\author{Long-Gang Pang$^{1, 2}$}
\email{lgpang.1984@berkeley.edu}

\author{Kai Zhou$^{4, 5}$}
\author{Xin-Nian Wang$^{1,2,3}$}

\address{$^{1}$Physics Department, University of California, Berkeley, CA 94720, USA}

\address{$^{2}$Nuclear Science Division, Lawrence Berkeley National Laboratory, Berkeley, CA 94720, USA}

\address{$^{3}$Key Laboratory of Quark \& Lepton Physics (MOE) and Institute of Particle Physics, Central China Normal University, Wuhan 430079, China}

\address{$^{4}$Frankfurt Institute for Advanced Studies, 60438 Frankfurt am Main, Germany}

\address{$^{5}$Institute for Theoretical Physics, Goethe University, 60438 Frankfurt am Main, Germany}

\begin{abstract}
    The structure of heavy nuclei is difficult to disentangle in high-energy heavy-ion collisions. The deep convolution neural network (DCNN)
    might be helpful in mapping the complex final states of heavy-ion collisions to the nuclear structure in the initial state.
    Using DCNN for supervised regression, we successfully extracted the magnitude of the nuclear deformation from event-by-event correlation
    between the momentum anisotropy or elliptic flow ($v_2$) and total number of charged hadrons ($dN_{\rm ch}/d\eta$) within a Monte Carlo model.
    Furthermore, a degeneracy is found in the correlation between collisions of prolate-prolate and oblate-oblate nuclei.
    Using the Regression Attention Mask algorithm which is designed to interpret what has been learned by DCNN,
    we discovered that the correlation in total-overlapped collisions is sensitive to only large nuclear deformation,
    while the correlation in semi-overlapped collisions is discriminative for all magnitudes of nuclear deformation.
    The method developed in this study can pave a way for exploration of other aspects of nuclear structure in heavy-ion collisions.
\end{abstract}

\keywords{Heavy ion collisions, Xe+Xe, U+U, deformed nuclei, machine learning}

\pacs{}

\maketitle

\section{Introduction}

The nuclear structure plays an important role in explaining the experimental data of heavy-ion collisions \cite{Heinz:2004ir,Kuhlman:2005ts,Filip:2009zz,Voloshin:2010ut,Goldschmidt:2015kpa,Alvioli:2011sk,Filip:2013mec,Denicol:2014ywa,Schenke:2014tga1,Adamczyk:2015obl,Alvioli:2018jls,Rybczynski:2017nrx,Noronha-Hostler:2019ytn}.
For example, experimental data from collisions of deformed uranium nuclei \cite{Adamczyk:2015obl} are found to favor one model of initial configurations, as described by semi-classical field of gluons from each nucleons \cite{McLerran:1993ni,McLerran:1993ka,Schenke:2012fw,Schenke:2014tga1}, whose initial entropy deposition does not have a distinctive linear dependence on the number of binary nucleon-nucleon collisions \cite{Miller:2007ri}. The mysterious enhancement of triangle flow in ultra central heavy-ion collisions can be partially resolved by considering many body quantum effects in the nuclear structure \cite{Alvioli:2011sk,Denicol:2014ywa,Alvioli:2018jls}. Despite of these empirical observations, a quantitative study of nuclear structure from high-energy heavy-ion collisions is still difficult because of the complexity of the final states.

Nuclear shape deformation is one aspect of the nuclear structure that can have observable influence on the hadron spectra and correlation in the final states of heavy-ion collisions.
A well established measurement of the nuclear shape deformation is the low energy Coulomb excitation \cite{MORINAGA1963210,CoulombExcitation1986}. When deformed nuclei pass through a thin slice of lead (Pb), some of the deformed nuclei are excited and deflected by the low energy Coulomb interaction. 
These excited nuclei radiate low-energy gamma rays that can be used to determine the nuclear shape deformation. The shape deformation of nuclear structure is used as input for the theoretical description of heavy ion collision \cite{Heinz:2004ir,Kuhlman:2005ts,Filip:2013mec}. It will be interesting to know whether the output of heavy-ion collisions is sufficient to constrain the nuclear shape deformation or other parameters in the nuclear structure despite of the highly complex and dynamical nature of the collisions. 

Mapping between two sets of data is always possible through deep neural network as long as there is a continuous geometric transformation \cite{Chollet:2017:DLP:3203489}. However, the power of mapping is not yet fully explored in regression tasks to map high dimensional scientific data to continuously changing control variables. If a brute force mapping using deep learning succeeds to build the connection, it can discover knowledge that may evade observation through conventional approaches. Such mapping can be made more efficient when the connection is already intuitively or evidently apparent. This is how a recent research was motivated where a deep learning system discovered the surprising connection between human gender and their retinal images \cite{Poplin2018}.

In this study we would like to use deep learning to map correlations between spectral observables to the initial nuclear deformation and explore whether the information on nuclear structure is encoded in the complex output of heavy-ion collisions using a Monte Carlo model.  If the connection exists, we will investigate whether the deep learning can decode this information from the output of heavy-ion collisions using supervised regression and understand what has been learned by the deep neural network.

\section{Results}

\begin{figure*}[!htp]
    \begin{center}
        \includegraphics[width=0.99\textwidth]{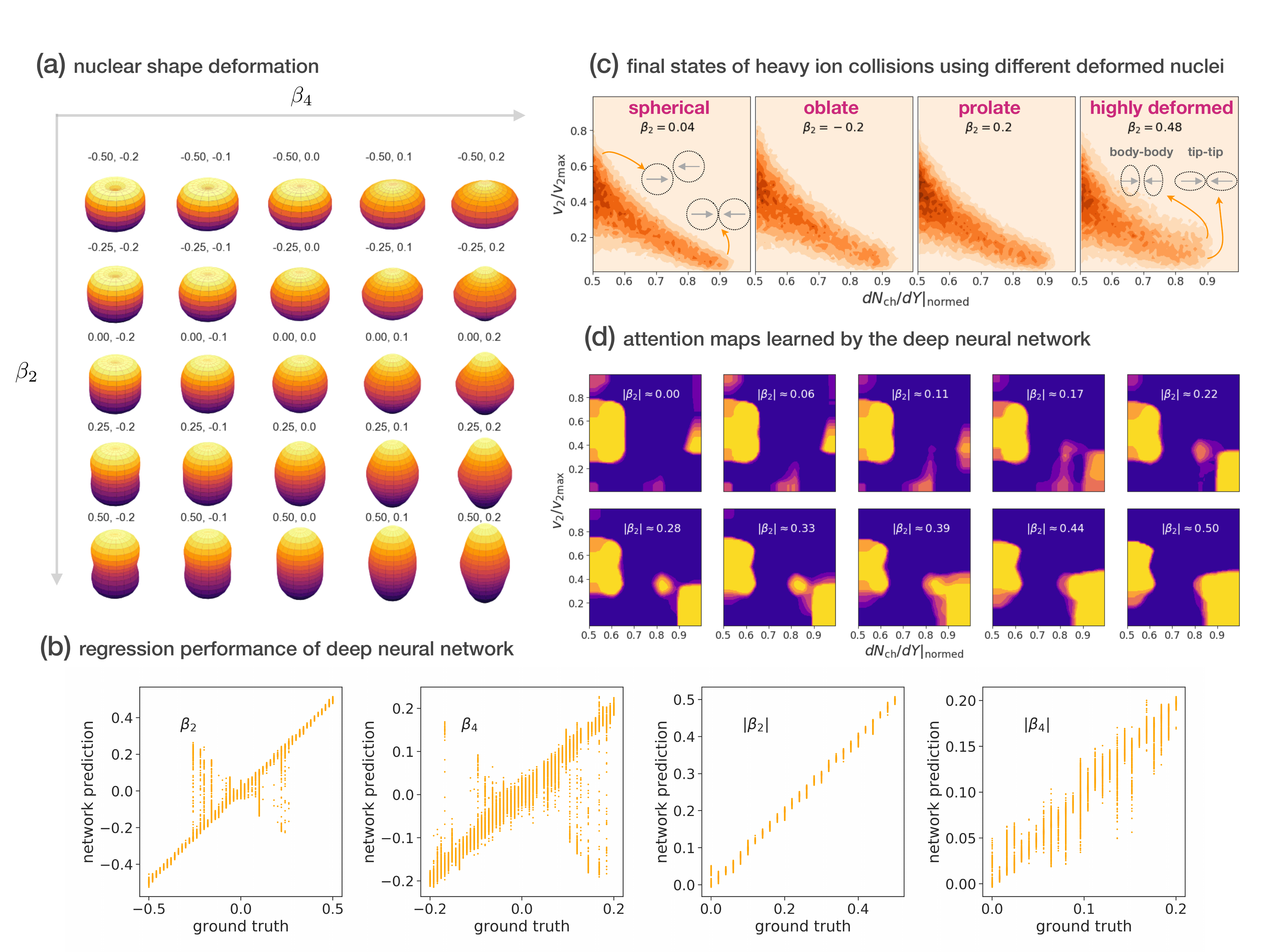}
    \end{center}
    \caption{(Color-online) Determining nuclear shape deformation using deep learning. (a) The three dimensional nuclear shapes as a function of two parameters 
    $\beta_2 \in [-0.5, 0.5]$ and $\beta_4 \in [-0.2, 0.2]$. (b) The regression performance of two deep convolution neural networks using the same architectures but different weights
    learned by setting labels to be $\beta_2$, $\beta_4$ (two left figures) and $|\beta_2|$ and $|\beta_4|$ (two right figures).
    (c) The complex Monte Carlo output for collisions of  different deformed nuclei. Deep learning uses these images as training and testing inputs.
    The x-axis represents the normalized total number of charged particles. The y-axis represents the normalized elliptic anisotropy of final state particles in momentum
    space.
    There is a degeneracy in the correlation between prolate and oblate nuclei as the network failed to predict the sign of the nuclear deformation.
    (d) The Regression Attention Mask helps to discover the most discriminative regions for nuclear deformation. While the ``ankle" region (semi-central collisions and large $v_2$) is sensitive to nuclear deformation, the ``toe'' region (central collisions and small $v_2$) is only sensitive to large nuclear deformation $|\beta_2|>0.17$.}
    \label{fig:dcnn_results}
\end{figure*}

The nucleon density distribution of deformed nuclei can be described by the deformed Woods-Saxon distribution. The deformation is controlled by two parameters, $\beta_2$ and $\beta_4$ (see Eq.~[\ref{woodssaxon}] in Section~\ref{method}), as visualized in Fig.~\ref{fig:dcnn_results}(a).
As the deformation parameter $\beta_2$ changes smoothly from negative to positive values,
the shapes of nuclei change from oblate (pumpkin-like) to prolate (egg-like).
We expect different patterns both in the initial energy density distribution and final state hadron spectra for collisions of different deformed nuclei.

To determine the nuclear shape deformation, we trained a deep convolution neural network (DCCN) to predict two deformation parameters $\beta_2$ and $\beta_4$ from physical observables obtained through theoretical simulations of heavy ion collisions using supervised regression as shown in Fig.~\ref{fig:dcnn_results}(b).  The physical observable we choose is the correlation between momentum anisotropy and total number of charged hadrons that is termed as charged multiplicity in the final state. The horizontal-axis represents the ground truth of deformation parameters and the vertical-axis represents the predictions by the deep convolution neural network. We observe that the predicted values by deep learning span a wide range from $-\beta_2$ to $\beta_2$ and $-\beta_4$ to $\beta_4$ in the left two sub-figures. The uncertainty range indicates that there is a degeneracy for the selected physical observable $ F(\beta_2) \approx F(-\beta_2)$. As a result, no inverse function $F^{-1}$  can map the selected physical observable to the sign of $\beta_2$. Knowing the degeneracy, we change our target to predict only the absolute values of the deformation parameters. This way, DCNN successfully extracted $|\beta_2|$ with small uncertainty and $|\beta_4|$ with medium uncertainty as shown in the right two sub-figures. The success in predicting the absolute values of the deformation parameters indicates that the nuclear deformation is encoded in the complex output of heavy ion collisions. The failure in extracting the sign of the parameters, on the other hand, indicates a degeneracy in the physical observable of the final state between prolate and oblate nuclei as discovered by the network. The statistical distributions of momentum anisotropy as a function of charged multiplicity verifies this degeneracy as shown in Fig.~\ref{fig:dcnn_results}(c),
where the training image for $\beta_2=-0.2$ is indistinguishable visually from  $\beta_2=0.2$. We will refer to the region of large multiplicity and small momentum anisotropy as the ``toe", and the medium multiplicity and large anisotropy as the ``ankle" in the statistical distributions of momentum anisotropy as a function of charged multiplicity in Fig.~\ref{fig:dcnn_results}(c).

To understand what has been learned by the deep neural network, we use ``Regression Attention Mask'' to highlight the most discriminative regions in the testing images as shown in Fig.~\ref{fig:dcnn_results}(d). We observed that the attention mask smoothly vary from spherical nuclei to highly deformed nuclei, indicating that the network has learned self-consistent features.

The most discriminative region is the ``toe" at the right bottom corner that corresponds to most central collisions. The ``Regression Attention Mask'' discovered this ``toe" region where the attention masks become higher and wider as $|\beta_2|$ goes from 0.17 to 0.5. The observation is consistent with physical intuition that fully overlapped tip-tip and body-body collisions of highly deformed nuclei have large momentum anisotropy fluctuations.

The ``toe" region discovered by deep learning has long been proposed to be sensitive to nuclear deformation. However, results from DCNN show that the ``toe'' is less sensitive to small deformations when $|\beta_2| < 0.17$ where the attention mask is very small. The regression mask also finds that the ``ankle" region for semi-overlapped collisions, where $dN_{\rm ch}/{dY}|_{\rm normed}\approx 0.5$, is sensitive to both small and large nuclear deformations.

\section{Discussion}

Our results suggest that the nuclear shape deformation is encoded in the complex outcome of heavy-ion collisions. Supervised regression in deep learning can decode part of the information from the final state outcome. DCNN can predict the deformation parameter $|\beta_2|$ to high accuracy. We have designed the Regression Attention Mask algorithm to locate important regions in the input image. The attention of the  artificial neural network vary smoothly as the value of $|\beta_2|$ increases. It does not only verify the old findings that fully overlapped collisions are sensitive to large nuclear deformation,
but also discovers new features in the region of semi-overlapped collisions, which work well to determine nuclear deformation both small and large.

The Regression Attention Mask is an important step towards the interpretable deep learning for science research. In the present study, the attention mask reveals interesting features that are also physically sound. For most central collisions, the attention mask finds the ``toe" region to be sensitive to large deformation, which corresponds to fully overlapped tip-tip and body-body collisions. For spherical nuclei with small $|\beta_2|$ on the other hand,  spatial eccentricity is strongly correlated with collision geometry with a thin ``toe". Attention mask suggests a large discriminative ``ankle" region for all values of  $|\beta_2|$,  because few events have extremely small or large $v_2$ in semi-overlapped collisions.

Much to our disappointment, deep learning fails to predict the sign of $\beta_2$ and $\beta_4$, indicating a degeneracy in the physical observable from collisions between prolate and oblate nuclei. Degeneracy can be observed directly in some nuclei, for example in Kr, whose ground-state wave function is a quantum superposition of prolate and oblate shapes \cite{Clement:2007zz}. The degeneracy we discover in the present study is with regard to observables in the final state of high-energy heavy-ion collisions. Data from heavy-ion experiments disfavor initial-state models whose entropy density deposition is linearly proportional to the number of binary collisions. As a result, tip-tip and body-body fully overlapped collisions produce similar numbers of charged particles and momentum anisotropy fluctuations for both prolate and oblate nuclei. It becomes clear when the 3-dimensional deformed nuclei are projected to 2-dimension by the extremely strong Lorentz contraction along the beam direction. The failure in predicting the sign of $\beta_2$ and $\beta_4$ using shallow and deep neural network indicates that the model is not over-fitting.

Such a degeneracy discovered by the network should not be surprising. If the physical process $F$ maps both $|\beta_2|$ and  $-|\beta_2|$ to the same final state observable $x$, it would be impossible for the network to find the inverse function $\beta_2 = F^{-1}(x)$. However, deep learning is helpful to efficiently verify the existence of an inverse function for the absolute value of $\beta_2$. In the present study, the network helps us to discover the existence of both the degeneracy and the inverse function $|\beta_2| = F^{-1}(x)$. %The same procedure is important for a wide range of regression problems in science research.
The sign of $\beta_2$ and  $\beta_4$ might be determined using data from other experiments such as low energy collisions or electron-ion collisions in which
%Using that data, it might be helpful to determine not only the nuclear shape, but 
one might be able to study other interesting nuclear structures such as the neutron skin, the electric charge and weak charge distribution, the pair correlation and the alpha clustering structure.

Our input images to the deep learning are the statistical information of engineered features. This is different from common computer vision problems where DCNN learns correlations between different patches of the same image. For scientific problems, the statistical information of many input samples from the same category is used to distinguish one category from another. It is also feasible to learn features in each event and use the statistical distribution of automatically learned features for the classification or regression task.

In the present study, we only use complete and semi-overlapped collisions where $v_2$ increases linearly as the initial state spatial eccentricity increases. For peripheral collisions where $dN_{\rm ch}/{dY}|_{\rm normed} < 0.5$, $v_2$ decreases as the spatial eccentricity continue to increase. The mapping function we used to get $v_2$ from spatial eccentricity does not work for peripheral collisions. It is the same reason for not using higher order momentum anisotropy as part of the training input.

A thorough study may require relativistic hydrodynamic simulations of heavy-ion collisions. The 3+1D hydrodynamic simulations may provide useful information that help to quantify the shape parameters, e.g., the event-plane twist along the longitudinal direction due to forward-backward asymmetry. This asymmetry not only arise in non-central collisions, but also in central (zero impact parameter) tip-body collisions. However, extending the present work to a full (3+1)D simulation is beyond our computational capability now. This might be feasible by running the recently developed GPU-parallelized hydrodynamic code in (2+1)D mode, e.g., CLVisc \cite{Pang:2018zzo} or GPU-VH \cite{Bazow:2016yra}.
In addition, one may improve the efficiency by selecting events with specific collision geometry provided that some regions are more discriminative in determining the nuclear shape deformation.

In summary, Monte Carlo simulations of heavy-ion collisions with various deformed nuclei reveal clear patterns in the complex final state, from which one can retrieve information about the structure of the initial state nuclei. Deep convolution neural network designed for classification is successfully used in regression task
to predict the magnitude of the nuclear deformation parameters from the correlation between momentum anisotropy and total hadron multiplicity. The network reveals that there is degeneracy between the outputs of prolate (positive $\beta_2$) and oblate (negative $\beta_2$) heavy-ion collisions. The Regression Attention Mask algorithm helps to locate the most discriminative regions in the input image.
It not only verifies that the DCNN learned the hidden structures which are sensitive to nuclear deformation, but also discovers a degeneracy in the sign of the nuclear deformation.

\section{Method}
\label{method}
Not all nuclei have a perfect spherical shape. Many nuclei have large deformations that lead to complex structures in the final state of heavy-ion collisions. For example, the collisions of prolate-shaped uranium nuclei have tip-tip, tip-body and body-body crossing patterns. The fluid dynamic expansion transfers the initial geometric eccentricity to the momentum anisotropy of the final state hadrons.
As a result, the most central tip-tip collisions have high multiplicity and small anisotropic flow while the body-body aligned collisions have similar multiplicity but large anisotropic flow for soft hadrons of low transverse momenta $p_T$. In this paper, we first train a 34-layer deep residual neural network \cite{Kaiming2015} with squeeze-excitation blocks \cite{squeezenet} to predict the shape deformation parameter of deformed nuclei using regression. Then we use the ``Regression Attention Mask '' to interpret what has been learned by the deep neural network.

\subsection{Collisions of deformed nuclei}

We use the Trento Monte Carlo model \cite{Moreland:2014oya} to provide IP-Glasma-like fluctuating initial conditions of heavy-ion collisions. The shapes of deformed nuclei  are given by the deformed Woods-Saxon distribution,
\begin{equation}
\rho(r, \theta, \phi) = \frac{\rho_0}{1 + e^{(r - R_0(1 + \beta_2 Y_{20}(\theta) + \beta_4 Y_{40}(\theta)))/a}}
\label{woodssaxon}
\end{equation}
where $\rho_0$ is the nucleon density in nucleus, $R_0$ is the Woods-Saxon radius, $\beta_2$ and $\beta_4$ are the deformation parameters introduced via an expansion in spherical harmonics,
$Y_{20} = \frac{\sqrt{5}}{4\sqrt{\pi}}(3\cos^2\theta - 1)$, $Y_{40} = \frac{3}{16 \sqrt{\pi}} (35 \cos^4 \theta - 30 \cos^2 \theta + 3)$ and $a$ is the Woods-Saxon tail width.

The orientations of the colliding nuclei are given by Euler rotations with random angles $(\alpha, \beta, \gamma)$.
\begin{equation}
    R(\alpha, \beta, \gamma) = R_z(\gamma) R_y(\beta) R_z(\alpha) 
    \label{eqn:euler_rotation}
\end{equation}
where $R_z(\alpha)$ is the first rotation along z-axis, $R_y(\beta)$ is the second rotation along y-axis and $R_z(\gamma)$ is the third rotation
along the original z-axis. Because the deformed nuclei are symmetric along the z-axis, the first rotation $R_z(\alpha)$ can be ignored.
To make sure the sampled rotations are isotropic, the tilt angle $\theta$ along y-axis is sampled according to a uniform distribution $\cos(\theta) \in U[-1, 1)$,
whereas the spin angle $\phi$ along z-axis is sampled according to a uniform distribution $\phi \in U[0, 2\pi)$.

We prepare 51x51=2601 groups of deformed uranium nucleus with 51 $\beta_2 \in [-0.5, 0.5]$ and 51 $\beta_4 \in [-0.2, 0.2]$. 
For each group, we simulate 100000 collisions with all possible collision geometries determined by the orientation of each nucleus and the impact parameter
(the transverse distance between the center of two colliding nuclei).
From these collisions we further select half of the events with highest total entropy,
which corresponds to centrality range $0-50\%$.

In experiments, the directly accessible information is the number of final state charged hadrons at mid-rapidity $dN_{\rm ch}/dY|_{Y=0}$
and the momentum anisotropy $v_2$ of final state hadrons.
It is shown in many studies that $dN_{\rm ch}/dY|_{Y=0}$ is proportional to the total entropy density $s_0$ of the initial state.
The anisotropy $v_2$ can be approximately computed from the geometric eccentricity of the initial state $\varepsilon_2 = \langle y^2 - x^2 \rangle /  \langle y^2 + x^2 \rangle$, where $x$ and $y$ are the transverse coordinates in the overlapped regions of collision, $\langle \cdots \rangle$ represents weighted average where weights are given by the local entropy density $s(x, y)$.
The geometric eccentricity in initial state transforms to momentum anisotropy in the final state through relativistic hydrodynamic expansion of the strongly coupled quark gluon plasma. To make the current method directly applicable to experiment, we match the $\varepsilon_2$ to $v_2$ through a heuristic equation \cite{Gardim:2011xv,Noronha-Hostler:2015dbi},
\begin{equation}
    v_2 = k_2 \varepsilon_2 + k_2' \varepsilon_2^3 + \delta_2
    \label{eq:ecc2_to_v2}
\end{equation}
where the coefficients $k_2=0.2$, $k_2' = 0.1$ and $\delta_2$ is the residual that  introduces additionally $\pm 10\%$ uniformly-distributed random fluctuations.

The total entropy is self-normalized with the mean entropy of the $0-1\%$ most central collisions for each nuclear shape deformation.
The self-normalization makes the method applicable to experimental data because 
\begin{equation}
    dN_{\rm ch}/{dY}|_{\rm normed} = \frac{dN_{\rm ch}/dY}{\langle dN_{\rm ch}/dY \rangle_{\rm 0\sim 1\%}} 
    \approx \frac{s_0}{\langle s_0 \rangle_{\rm 0-1\%}} .
    \label{eq:s0_scale}
\end{equation}

We now have 2601 groups of $(dN_{\rm ch}/dY|_{\rm normed}, v_2)$ distributions.
The data are divided into 3 groups, $80\%$ for training, $10\%$ for validating and $10\%$ for testing.
We use data augmentation to enlarge the size of the training data set. For each distribution, we randomly sample $90\%$ from 50000 data points to create a new image.
The data augmentation produces $160000$ images for training, $16000$ for validating and $16000$ for testing.

\subsection{Deep regression network}

\begin{figure}[!htp]
    \begin{center}
        \includegraphics[width=0.5\textwidth]{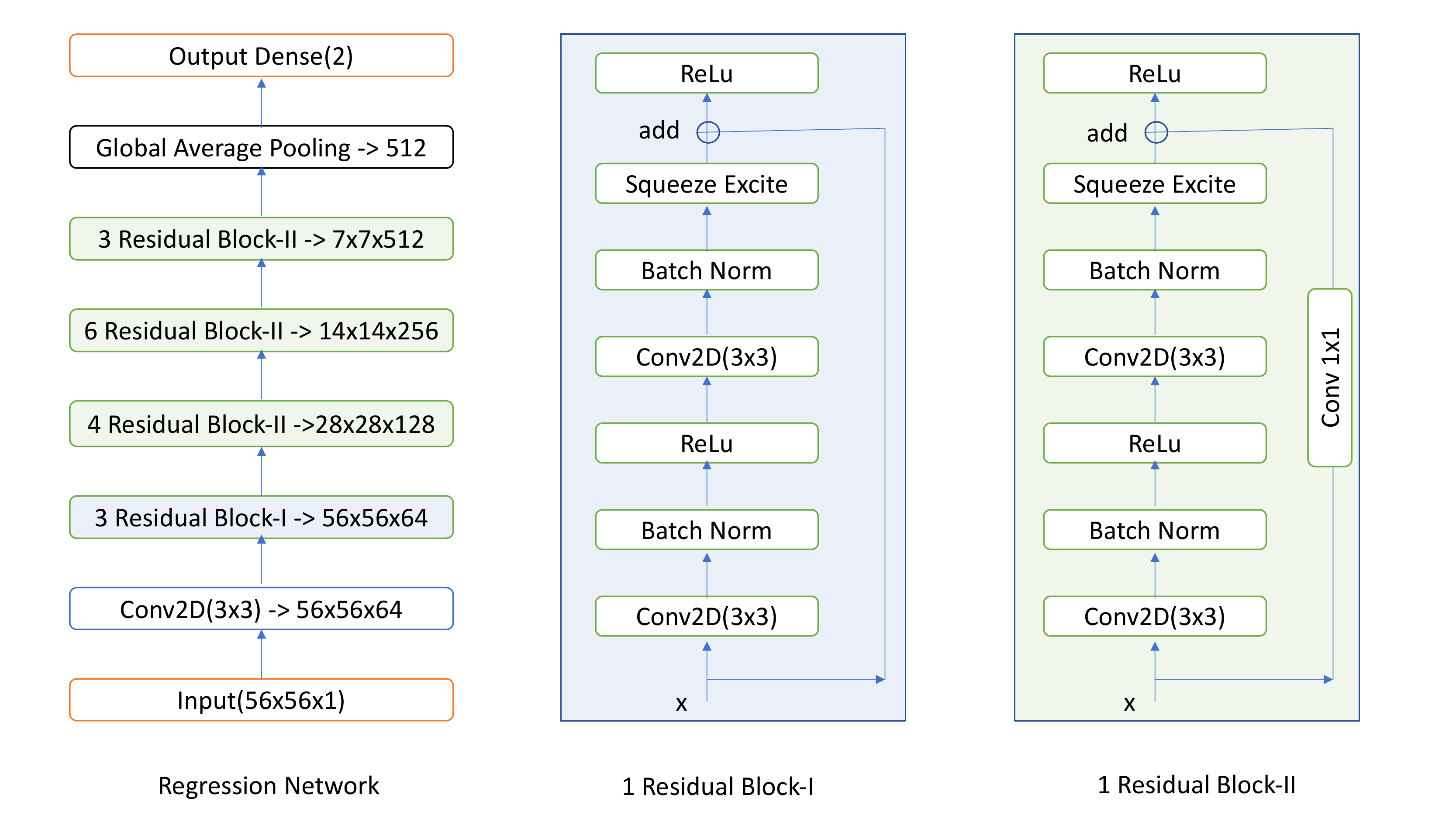}
    \end{center}
    \caption{(Color-online) The architecture of the 34-layer regression neural network using residual and squeeze excitation blocks.}
    \label{fig:network}
\end{figure}

Shown in Fig.~\ref{fig:network} is the 34-layer deep convolution neural network for the regression task.
The residual blocks make it possible to design deep convolution neural network for image classifications.
And the squeeze excitation operation additionally pushes the image classification to the state-of-the-art.
We verify that the deep residual neural network designed for image classification also works well on regression task.
Our inputs are images of 2 dimensional event-by-event distributions of $(dN_{\rm ch}/dY|_{\rm normed}, v_2/v_{2 \rm max})$ in $56\times56$ bins.
The input image is first processed using a two-dimensional convolution, then it is fed to a type-I residual box containing 3 blocks named Residual Block-I,
where the output feature maps have the same transverse size as the input image.
The resulted feature maps are fed to four type-II residual boxes consecutively.
Each type-II residual box has $3$ to $6$ blocks named Residual Block-II.
The first Residual Block-II in each box reduces the width and height of the input feature map by a factor of 2.
All the residual blocks have one ``add'' operation and the last ``add'' layer has a name ``add\_16''.
Each residual block has 2 Conv2D layers and in total they contribute to $16 \times 2 = 32$ convolution layers.
We have used global average pooling layer \cite{saliency_maps} to get the mean of each feature map with size $7\times7$ for the $512$ channels.
This $512$ neurons are connected to $2$ neurons in the output layer to make predictions 
for the nuclear deformation parameter $|\beta_2|$ and $|\beta_4|$.
One reason to use this deep residual neural network is to verify whether a deep network can learn
the sign of the parameter $\beta_2$ where a shallow network has already failed. The residual neural network also has better interpretability than VGG-like network as shown in the paper \cite{netdissect2017}.

\subsection{Regression Attention Mask for interpretable deep learning}

Interpretability is the most indispensable consideration of the deep neural networks
when it is used in science researches, as well as self-driving cars, medical diagnosis and government policy making. The interpretability is defined as the {\it ability to explain or to present in understandable terms to a human} \cite{Doshi2017}. Visualization, verbal explanation and clustering of similar instances are all understandable representations of the deep neural networks with Interpretability.

There are many ways to visualize what has been learned by the network classifier. For reviews see the book ``Interpretable Machine Learning'' \cite{Christoph2019} and surveys \cite{Zhang2018,survey_for_explaining,XAI,Doshi2017}.
There are global explanations that explain the network in the whole input space by visualizing what each feature map learns. There are also local explanations that explain local features in one specific image. We have designed the ``Regression Attention Mask'' algorithm, which provides a local interpretation of a given image.

For the global explanation, deconvolution is used to visualize each feature map of the deep convolution neural network \cite{Erhan:2009,deconv2013,olah2017feature,olah2018the}. 
For the local explanation, there are many different methods developed based on the assumption that one highly complex machine learning model can be locally approximated by a linear model around one given input image. One way to construct the importance map is to measure the probability changes with parts of the image occluded \cite{occlude, prediction_difference_analysis} or similar pixels/super-pixels (LIME) masked \cite{lime}.
Different from our method, those occlusion methods depend on manually constructed masks of input images.

Saliency map is another way to explain the pre-trained convolution neural network around one given image \cite{saliency_maps}. It assumes that the predicted class score can be approximated by a linear function $f(x) \approx w \circ x + b$ around one given image $x$ in the input space, where $f$ is the function learned by the network.
The gradient $w = \partial f / \partial x$ represents the importance of each pixel.
However, the original saliency map is noisy \cite{saliency_maps} due to negative gradients and non-linear dependences on $x$.
The improved saliency map uses guided back-propagation \cite{guided_backprop}
to maximize the class score of one given class by dropping negative influences.
These gradient based methods as well as many alternatives \cite{integrated_grad,visual_backprop,multicam}
are sensitive to constant shift \cite{unreliable_saliency} except the pattern net \cite{deep_taylor,pattern_net}.
The interpretability of all different saliency maps can be quantified using our ``Regression Attention Mask '' method.

What is closely related to our method is the class activation map where locations in the feature maps of the last convolution layer are matched to the input image \cite{zhou2015cnnlocalization,gradcam}. We have discarded the RELU activation function in the gradient weighted class activation map to get our specific activation map for regression tasks. RELU pick positive influence to enlarge the class score while regression needs both positive and negative components in the activation map to reproduce the regression value. The regression activation map is used to create ``Regression Attention Mask''.

Based on the class activation map, the class activation mask  is invented to quantify the interpretability of different neural networks. The class activation mask is a two-dimensional image that has the same size as the input image.
It has only one channel and its pixel values are initialized with $0$.
Pixels are set to $1$ if corresponding regions in class activation map have values larger than some threshold. The interpretability of one classifier is quantified by the intersection over union score between the class activation mask and 
human understandable concept-segmentation, e.g., human labeled masks for an object, part, scene, material, texture and color \cite{netdissect2017}.
The interpretability has the order ResNet > VGG > GoogLeNet > AlexNet regarding different network architectures.  Different from that method, we propose to use prediction difference of the masked image to quantify the interpretability in the regression network.

To disentangle hidden representations of the learned feature maps, studies in Refs. \cite{Zhang_2018_AAAI} and \cite{Zhang_2018_CVPR} use graphs, decision trees and local part template. Recently a deep neural network has been trained to jointly classify images into categories and provide its reasoning \cite{verbal}. Our framework provides an explanation about its decisions in the regression task and helps us to understand the features of the correlation in determination of the nuclear deformation.

For classification task, the importance of each pixel to classification can be computed using the gradient weighted class activation map (Grad-CAM),
\begin{equation}
    gradcam(x) = \frac{1}{c\times k \times k} \sum_{n=1}^{c} A^n \sum_{i,j=1}^{k} \frac{\partial f}{\partial A_{ij}^n}
    \label{eq:grad_cam}
\end{equation}
where $x$ is the input image, $c$ is the number of channels, $k$ is the size of the activation map, $f$ is the class score, $A_{ij}^n$ is the pixel value of the
$n$th activation map $A^n$ in layer ``add\_16'' at site $(i,\ j)$.
The class activation map is scaled up to the same size as the input image by upsampling. In the original grad-cam paper, the weighted class activation map is forwarded to a ReLU activation function to remove negative contributions.
Otherwise the positive influence on one class might be equally negative on the other to cancel the important regions, when the prediction probabilities are close for the top-2 classes. However, both positive and negative contributions are required to reproduce the regression value. Different from the original grad-cam algorithm, the ReLU activation function in our algorithm is removed to adapt to the regression task.

The attention mask for input image $x_i$ is defined as $m_i = gradcam(x_i) > T$ where $T$ is the threshold.
In the present study, the threshold $T$ is set to the mean value of the given mask.
Since the input images have similar structure for the same $\beta_2$ and $\beta_4$, we compute the averaged attention mask $m = \sum_i w_i m_i$, 
for all events in the range $ \in [|\beta_2|, |\beta_2|+0.02]$, weighted by $w_i$,
\begin{equation}
    w_i = \frac{\exp\left[ -\sigma_i \right]}{\sum_{j} \exp\left[ -\sigma_j\right]}, \quad\; \sigma_i = ||f(m_i \circ x_i) - f(x_i)||,
    \label{eq:cam_weights}
\end{equation}
where $x_i$ is the $i$th input image, $m_i$ is the attention mask of the trained regression network.
The $m_i \circ x_i$ is the pixel-wise multiplication between the attention mask and the input image,
which helps to occlude unimportant regions.
Feeding the original image $x_i$ and the occluded image $m_i \circ x_i$ to the regression network $f$ helps to
get the prediction difference $\sigma_i$. 
Smaller prediction difference indicates better attention mask that leads to higher weight $w_i$.

\begin{acknowledgments}
We thank Volker Koch, Jorgen Randrup, Feng Yuan and Xin Dong for helpful discussions.
This work is supported by DOE under Contract No. DE-AC02-05CH11231, by NSF under Grant No. ACI-1550228 within the JETSCAPE Collaboration, by NSFC under Grant No. 11861131009 and No. 11890714,
by BMBF under the ErUM-Data project and the AI research grant of SAMSON AG, Frankfurt.
Computations are performed on GPU workstations at CCNU and DOE NERSC.
\end{acknowledgments}

\bibliographystyle{unsrt}

\bibliography{inspire,not_inspire}

\end{document}